\documentstyle[twocolumn,aps,epsf]{revtex}
\begin{document}
\title{
Imaging the Phase of an Evolving Bose-Einstein Condensate Wavefunction}
\author{
J. E. Simsarian,$^{1}$ 
J.\ Denschlag,$^{1}$
Mark Edwards,$^{1,2}$
Charles W. Clark,$^{1}$ 
L.\ Deng,$^{1}$
E. W.\ Hagley,$^{1}$
K.\ Helmerson,$^{1}$
\newline   
S. L.\ Rolston,$^{1}$ and W.D.\ Phillips$^{1}$
}
\address{
$^{1}$National Institute of Standards and Technology, 
Gaithersburg, MD
20899\\
$^{2}$Georgia Southern University, Statesboro, GA 30460-8031}
\date{\today}
\maketitle
\begin{abstract}
We demonstrate a spatially resolved autocorrelation measurement with a 
Bose-Einstein condensate (BEC) and measure the evolution of the 
spatial profile of its quantum mechanical phase.  Upon release of the BEC 
from the magnetic trap, its phase develops a form that we measure to be 
quadratic in the spatial coordinate.  Our experiments also reveal the effects
of the repulsive  interaction between two overlapping BEC wavepackets and we
measure the  small momentum they impart to each other.

\end{abstract}

A trapped Bose-Einstein condensate~\cite{anderson0} has unique 
value as a source for atom lasers~\cite{mewes} and matter-wave 
interferometry~\cite{berman} because its atoms occupy the same 
quantum state, with uniform spatial phase.  However, when released from the trapping 
potential, a BEC with repulsive atom-atom interactions expands, 
developing a non-uniform phase profile.  Understanding this phase 
evolution will be important for applications of coherent matter waves.  
We have developed a new interferometric technique using spatially 
resolved autocorrelation to measure the functional form and time 
evolution of the phase of a BEC wavepacket expanding under the influence of its 
mean field repulsion.

In 1997, the coherence of weakly 
interacting BECs was demonstrated by releasing two spatially separated condensates and 
observing their interference
\cite{andrews}.  Subsequent experiments have further investigated
condensate coherence properties. One~\cite{stenger} 
used velocity-resolved Bragg diffraction
\cite{kozuma} to probe the momentum spectrum of
trapped and released BECs.  A complementary
experiment~\cite{hagley2} that used matter-wave interferometry can be 
interpreted as a measurement of the spatial correlation function, 
whose Fourier transform is the momentum spectrum.
These experiments showed that a trapped condensate has a
uniform phase, and a released condensate develops a non-uniform
phase profile.  (Recently the influence of non-zero temperature 
on coherence 
properties was also investigated~\cite{bloch}).  The
experiments reported in this Letter combine spatial resolution and 
interferometry to measure the functional form of the time-dependent 
phase profile of a released condensate.  We also make the first 
measurement of the velocity imparted to two equal BEC 
wavepackets from their mutual mean-field repulsion~\cite{ketterle}.
 
We perform our experiments with a condensate of $1.8(4) \times 
10^6$~\cite{uncertainty}
sodium atoms in the $3S_{1/2}$, $F=1$, $m_{F}=-1$ state.  The sample has
no discernable non-condensed (i.e. thermal) component.
The condensate is prepared following the method of Ref.~\cite{kozuma}
and is held in a magnetic trap
with trapping frequencies $\omega_x  =  \sqrt{2}\omega_y = 2 \omega_z 
= 2\pi \times $27 Hz. Using a scattering length of $a = 2.8$ nm, the 
calculated Thomas-Fermi diameters~\cite{dalfovo} are 47 $\mu$m,
66 $\mu$m, and 94 $\mu$m, respectively.

We release the BEC from the magnetic trap and it expands, driven mostly by 
the mean-field repulsion of the atoms.  This expansion implies the 
development of a nonuniform spatial phase profile (recall that the 
velocity field is proportional to the gradient of the quantum 
phase).  
After an expansion time $T_{0}$, we probe the phase profile
with matter-wave Bragg interferometry~\cite{giltner,torii,denschlag}.
Our interferometer splits the BEC into two wavepackets and
recombines them with a chosen overlap, producing interference 
fringes, which we measure with absorption imaging~\cite{imaging}.
From the dependence of the fringe spacing on the overlap, we extract 
the phase profile
of the wavepackets.
 
\begin{figure}
\leavevmode
\centering
\epsfxsize=3.2in
\epsffile{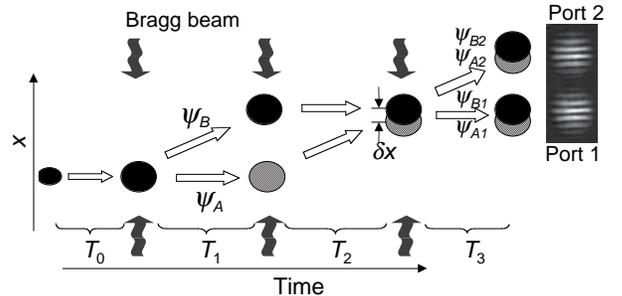}
\caption{
Space-time diagram of the experiment.  Three optically induced 
Bragg-diffraction pulses form the interferometer.  The condensate is 
released for a time $T_{0}$ before the first Bragg pulse.  
The centers of $\psi_{{A}}$ and $\psi_{{B}}$
are separated by $\delta x$ at the time of the third Bragg pulse,
which splits them into  $\psi_{{A1}}$,
$\psi_{{B1}}$, and  $\psi_{{A2}}$,  $\psi_{{B2}}$.
Before imaging the atoms, we allow the output ports to separate for a 
time $T_{3} \approx 2$ ms.  The
image shows the output ports when $T_{0}$ = 3 ms, 
$T_{1}$ = 1 ms, and $T_{2}$ = 1.3 ms.
\label{fig1}}
\end{figure}

Our atom interferometer~\cite{denschlag} consists of
three optically-induced Bragg-diffraction pulses  
applied successively in time (Fig.\ 1).
Each pulse consists of
two counter-propagating laser beams  whose frequencies differ
by 100 kHz. They are detuned by about $-2$ GHz
from atomic resonance ($\lambda = 2\pi/k = 589$ nm) so that spontaneous
emission is  negligible.  The first pulse has a duration of 6 $\mu$s and 
intensity sufficient to provide a $\pi/2$ pulse, which coherently splits the BEC
into two wavepackets, $\psi_{{ A}}$ and $\psi_{{
B}}$.  The wavepackets have about the same number of atoms and only
differ in their momenta: $p=0$ and $p=2\hbar k$. At a time $T_{1}=1$ ms after the first 
Bragg pulse, the two wavepackets are completely separated and
a second Bragg pulse (a $\pi$ pulse) of 12 $\mu$s  duration transfers
$\psi_{{ B}}$ to a state with $p\approx 0$ and $\psi_{{ 
A}}$ to $p\approx 2\hbar k$~\cite{almost}.
After a variable time $T_{2}$ the wavepackets partially overlap again
and we apply a third pulse, of
6 $\mu$s  duration (a $\pi/2$ pulse).
This last pulse splits each wavepacket into the two momentum states.  The 
interference of the overlapping wavepackets in each of the two momentum 
states allows the determination of the local phase difference between them.
By changing the time $T_{2}$ we vary $\delta x = x_{ A} - x_{ B}$, 
the separation of $\psi_{{A}}$ and $\psi_{{B}}$ 
at the time of the final Bragg pulse.
The set of data at different $\delta x$ constitutes a new type of 
spatial autocorrelation
measurement that is similar to the ``FROG'' technique~\cite{frog} used to measure the 
complete field of ultrafast laser pulses.
From these measurements we obtain the phase profile of the 
wavepackets in the $x$ direction.

\begin{figure}
\leavevmode
\centering
\epsfxsize=3.3in
\epsffile{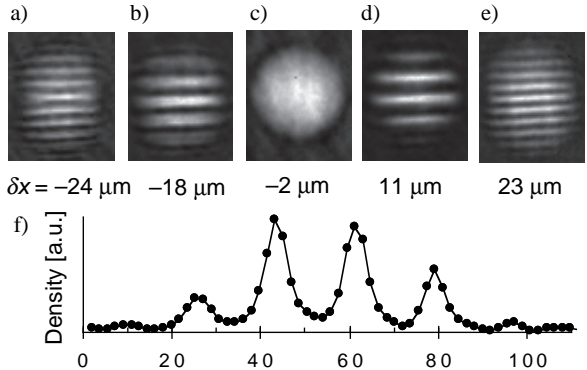}
\caption{
(a-e) One of the two output ports of the interferometer with $T_{0}$ =
4 ms and $\delta x$ as indicated.  (f) A plot of the density along the $x$
direction of (d).  
\label{fig2}}
\end{figure}

Figure \ref{fig2}a-e shows one interferometer output port for
different $\delta x$ (different $T_2$) after an expansion
time $T_{0}$ = 4 ms. In general, we observe straight, 
evenly spaced fringes (although for small $T_{0}$ 
and $T_{2}$ the fringes may be somewhat curved).  There is a value of
$\delta x = x_{0} \neq 0$ where we observe no fringes (Fig.\ \ref{fig2}c)  
and the fringe spacing decreases as $|\delta x - x_{0}|$ increases.
Figure \ref{fig2}f, a cut through Fig.\ \ref{fig2}d, shows the 
high-contrast fringes~\cite{contrast}.
Our data analysis uses the average fringe period $d$, obtained from plots 
like Fig.\ \ref{fig2}f.

The fringes come from two different effects:  the
interference of two wavepackets with quadratic phase profile, and
a relative velocity between the wavepackets' centers.
The data can be understood by calculating the fringe spacing along $x$ 
at output port 1~\cite {separable}.
We assume that the phase $\phi$ of the wavefunction $f {\rm
e}^{i\phi}$ can be written
as $\phi = \frac{\alpha}{2}x^{2}+\beta x$.  The equal spacing of the 
fringes implies, as predicted in the Thomas-Fermi limit 
\cite{castin}, that $\phi$ has no significant higher-order terms 
\cite{polynomials}.  The curvature coefficient $\alpha$
describes the mean-field expansion of the wavepackets and $\beta$
describes a relative repulsion velocity.  The velocity arises 
because the wavepackets experience a repulsive push as they first 
separate and again as they recombine.
The density at port 1 (see Fig.\ 1) just after the final
interferometer pulse is the interference pattern 
$|{\psi_{A1} + \psi_{B1}}|^{2}$ of the wavepackets $\psi_{{A1}}$ 
and $\psi_{{B1}}$:
\begin{equation}
|f(x-\delta x)e^{i(\frac{\alpha}{2}(x-\delta x)^{2}
- \beta(x-\delta x))} +  f (x)e^{i(\frac{\alpha}{2} x^{2} +
\beta x)}|^{2},
\label{wavefunctions}
\end{equation}
where we assume that the amplitudes and curvatures of the wavepackets are 
equal and their velocities have equal magnitude and opposite direction. The cross term of 
(\ref{wavefunctions}) is
\begin{equation}
2f(x-\delta x)f(x){\rm cos}\left[ \left(\alpha \, \delta x +
\frac{{M}\, \delta v}{\hbar}\right)x + C\right],
\label{fringes}
\end{equation}
where $M$ is the sodium mass, ${M}\, \delta v/\hbar \equiv
2 \beta$, and $C$ is independent of $x$~\cite{constant}.  
$\delta v = v_{{B}} - v_{{A}}$ is the relative repulsion velocity
between the wavepackets $\psi_{{A1}}$ 
and $\psi_{{B1}}$. Expression (\ref{fringes}) predicts
fringes with spatial frequency,
\begin{equation}
\kappa  = \alpha \, \delta x + \frac{{M}\, \delta v}{\hbar}, 
\label{line}
\end{equation}
where $|\kappa| = 2 \pi/d$.  When there are no fringes, 
$\kappa$ = 0 and the wavepacket separation $\delta x =
x_{0} \equiv  -{M}\, \delta v/ \alpha \hbar$.

\begin{figure}
\leavevmode
\centering
\epsfxsize=3.3in
\epsffile{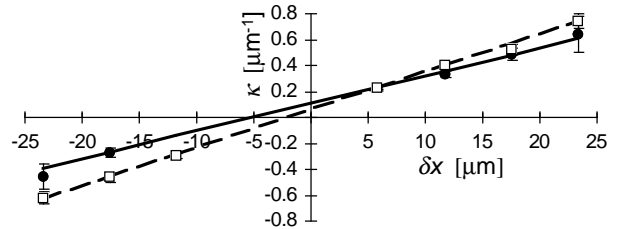}
\caption{
Plot of the spatial fringe frequency $\kappa$ versus $\delta x$ for $T_{0}$ = 1 ms (filled 
circles) and 4
ms (open squares).  The solid and dashed lines are linear fits to the data.  
\label{fig3}}
\end{figure}

Figure \ref{fig3} plots the measured
$\kappa $ vs. $\delta x$~\cite{separation} for $T_{0}$ = 1 and 4 ms.
The data are well fit by a
straight line as expected from Eq.\ (3) in the approximation that 
$\alpha$ and $\delta v$ are independent of $\delta x$.  The slopes of 
the lines are the phase curvatures 
$\alpha$, and the $\kappa$ intercepts give the relative velocities 
$\delta v$.

We checked the validity of the data analysis procedure by 
analyzing data simulated with a 1-D Gross-Pitaevskii (GP) treatment.  
Despite variations of $\delta v$ and $\alpha$ with $\delta x$ (due 
to their continued evolution during the variable time $T_{2}$), we find 
that $\kappa$ is still linear in $\delta x$.  The slopes and intercepts in general
are averages over the range of $\delta x$ used
in the experiment.

The interference fringes used to determine $\alpha$ and
$\delta v$ are created at the time of the final interferometer
pulse.  Because the two outputs overlap at that moment, we wait 
a time $T_{3}$ for them to separate before imaging.  During this time, 
the wavepackets continue to expand.
The 1-D simulations show that the fringe 
spacings and the wavepackets expand in the same proportion.  We correct
$\kappa$ (by typically 15 $\%$) for this, using the calculated expansion
from a 3-D solution of the GP equation described below.

\begin{figure}
\leavevmode
\centering
\epsfxsize=3.3in
\epsffile{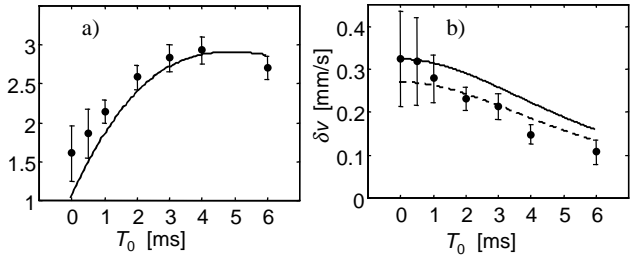}
\caption{
(a) Plot of the phase curvature $\alpha$ versus the initial expansion time
$T_{0}$ showing  the phase evolution from mean-field expansion.  The solid
line  is a calculation using the Lagrangian Variational Method (LVM).
(b) Plot of the relative repulsion velocity $\delta v$ versus $T_{0}$.  The 
solid curve is the calculated maximum repulsion velocity (when $\delta x = 0$) and the 
dashed curve is the
repulsion velocity averaged over the range of $\delta x$ used in the
experiment. 
\label{fig4}}
\end{figure}

The different slopes and intercepts of the two lines in Fig.\ 
\ref{fig3} show that the curvature
$\alpha$ and relative velocity $\delta v$ of the wavepackets depend
on the release time $T_{0}$ before the first interferometer pulse.
Figure \ref{fig4} plots the dependence of
$\alpha$ and $\delta v$ on various release times $T_{0}$.
The condensate initially has a uniform phase so that immediately
after its release from the trap $\alpha = 0$.
We nevertheless measure a nonzero $\alpha$ for $T_{0}$ = 0 ms because the BEC 
expands during $T_{1}$ and $T_{2}$.  As a function of time, $\alpha$
behaves as $\dot D$/$D$ where $D$ is the wavepacket diameter and $\dot D$ is its 
rate of change~\cite{castin}.  At early times when the mean-field energy is 
being converted to kinetic energy, $\dot D$
increases rapidly, {\em increasing} $\alpha$.  At late
times, after the mean-field energy has been converted, 
$D$ increases while $\dot D$ is nearly constant, {\em decreasing} $\alpha$.

We predict the time evolution of $\alpha$ using the Lagrangian
Variational Method (LVM)~\cite{perez}.  The LVM
uses trial wavefunctions with time dependent parameters to provide
approximate solutions of the 3-D time-dependent GP 
equation. In the model, the effect of the interferometer pulses is to 
replace the original wavepacket with a superposition of
wavepackets having different momenta; e.g., the action of our first 
interferometer pulse is 
$\psi_{0} \rightarrow \left(\psi_{0} + e^{i2kx}
\psi_{0}\right)/\sqrt{2}$.  We use Gaussian trial wavefunctions
in the LVM and, for simplicity, neglect the interaction between the
wavepackets, to  calculate the phase curvature $\alpha$ at the time of the
last interferometer pulse.  This result, with $T_{1} = T_{2}$,
is the solid line of Fig.\ 4a.

We use energy conservation to calculate the relative repulsion velocity 
$\delta v$ between $\psi_{A1}$ and $\psi_{B1}$ because we neglect wavepacket
interactions in the LVM.  
In the Thomas-Fermi approximation, we can calculate the amount of energy
available for repulsion when $T_0$ = 0.  
A trapped condensate has $\frac{5}{7}\mu$ average
total energy per particle, where $\mu$ is the chemical
potential~\cite{dalfovo}.  After release from the trap, it has $\frac{2}{7}
\mu$ average mean-field energy per particle.  Applying a $\pi/2$ Bragg pulse
to the BEC causes a density corrugation, which increases
the mean-field energy to $\frac{3}{7}\mu$ per particle.  In the
approximation that the wavepackets do not deform as they separate and
recombine, one can show that 1/3 of the total mean-field energy goes
into expansion of the wavepackets, and 2/3 is available 
for kinetic energy of center-of-mass motion.  Therefore 
$\frac{2}{7}\mu$ of mean-field energy per particle is available for
repulsion.  The  corresponding repulsion velocity is only about $10^{-2}$ of a
photon recoil velocity.  The repulsion energy and $\delta v$ decrease for
larger $T_0$ because both are inversely proportional to the condensate volume,
which we calculate with the LVM.  The two curves shown in 
Fig.\ \ref{fig4}b are the calculated $\delta v$ when $\delta x = 0$ 
(solid curve) and $\delta v$ averaged over the different $\delta x$ used in 
the experiment (dashed curve).  The 1-D GP simulations suggest that for small 
$T_{0}$, the results of the experiment should be closer to the solid curve;
and for large $T_0$, closer to the dashed curve.  The data is  consistent with
this trend.

\begin{figure}
\leavevmode
\centering
\epsfxsize=3.4in
\epsffile{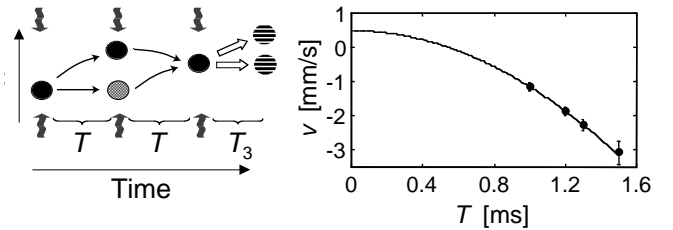}
\caption{
(a) Schematic representation of the interferometer in 
the trap, with the principle difference from Fig.\  1 being the curved
arrows indicating the acceleration of the wavepackets. (b) The relative 
velocity $v$ between the two trapped wavepackets versus the
interferometer time $T$.  The solid line is a fit.   
\label{fig5}}
\end{figure}

In a related set of experiments we performed interferometry in the
trap.  This differs from the experiments on a released BEC because 
there is no expansion before the first interferometer 
pulse~\cite{contraction} and the
magnetic trap changes the relative velocity of the
wavepackets between the interferometer pulses (Fig.\ 5a).
To better reveal the velocity differences, we choose $T_{1}$ =
$T_{2} = T$ to suppress fringes arising from the phase curvature.
As with the released BEC measurements, we observe equally spaced 
fringes at the output of the interferometer,
although the fringes are almost entirely due to a relative velocity $v$ between
the wavepackets $\psi_{{A1}}$ and $\psi_{{B1}}$ at the time
of the third interferometer pulse.
We obtain $v$ from the fringe periodicity after a small correction 
for residual phase curvature~\cite{velcorrection}.

Two effects contribute to $v$: the mutual repulsion
between the wavepackets $\psi_{{A}}$ and $\psi_{{B}}$
and the different action of the trapping potential
on the two wavepackets in the interferometer.
The latter effect occurs because after the first Bragg pulse, $\psi_{{A}}$ remains
at the minimum of the magnetic potential while
$\psi_{{B}}$ is displaced.
Wavepacket $\psi_{{B}}$ therefore spends more time away from the 
center of the trap and experiences more acceleration than $\psi_{{A}}$.

Following the last Bragg pulse,
$\psi_{{A1}}$ and $\psi_{{B1}}$ have a velocity difference
which for our parameters can be approximated by
$v \approx -\frac{2\hbar k}{M}{\rm sin^{2}}(\omega_{x} {T}) + \delta v$~\cite{approx}. 
Figure \ref{fig5}b plots $v$ versus $T$, and the curve
is a fit to the above expression.  We obtain the trap frequency 
$\omega_x /2\pi = 26.7(15)$ Hz, in excellent agreement with an independent 
measurement.  We also obtain the relative
velocity from the mean-field repulsion $\delta v = 0.49(12)$ mm/s, 
which we expect to be somewhat larger than for the released 
measurements because the wavepackets contract, producing a 
larger mean field.

In conclusion, we demonstrate an autocorrelating matter-wave
interferometer and use it to study the evolution
of a BEC phase profile by analyzing spatial images of interference 
patterns.  We study how the phase
curvature of the condensate develops in time and measure the repulsion
velocity between two BEC wavepackets.  Our interferometric method 
should be useful
for characterizing other interesting condensate phase profiles. 
For example, it can be applied to detect excitations
of a BEC with characteristic phase patterns,
such as vortices and solitons
\cite{denschlag,burger,matthews,madison,jackson}.
The method should be useful for further
studies of the interaction of coherent wavepackets and to study the
coherence of atom lasers.

We thank T. Busch, D. Feder, and L. Collins for helpful discussions. This work
was supported in part by the US Office of Naval Research and NASA.
J.D. acknowledges support from the Alexander
von Humboldt foundation.  M.E. and C.W.C. acknowledge partial support from NSF 
grant numbers 9802547 and 9803377.


\begin{references}
\bibitem {ketterle} The mean-field 
energy shift of the out-coupled wavepacket in~\cite{stenger} can be used to
infer the repulsion velocity.  Here we measure the
velocity directly. %
\bibitem {uncertainty} All uncertainties reported here are 1 standard
deviation combined statistical and systematic uncertainties.
\bibitem {imaging} The condensate was imaged by
first optically pumping the atoms to the $F=2$ ground
state and then imaging the absorption
of a probe beam on the $F = 2 \rightarrow F' = 3$
transition.  The pulse had a 5 $\mu$s duration, $\approx$
170 mW/cm$^{2}$ intensity, and was detuned 15 MHz from resonance.
\bibitem {almost} The momenta are not exactly $p=0$ and $p=
2 \hbar k$ because of repulsion effects that will be discussed.
\bibitem {contrast} The observation is consistent with the full 
predicted fringe contrast when we include the finite imaging resolution.
\bibitem {separable} We can treat the $x$ direction independently of 
$y$ and $z$ when the wavefunction $\psi= f e^{i\phi(x,y,z)}$ is separable, i.e.
$\phi(x,y,z) = \varphi(x) +  \eta(y,z)$.  Straight fringes imply this 
separability.
\bibitem {polynomials} For $T_{0} =
6$ ms and an interference pattern with many fringes, 
we found that the coefficients of third and fourth order terms are smaller than 1$\times 10^{-5}$
$\mu$m$^{-3}$ and 1$\times 10^{-7}$ $\mu$m$^{-4}$ respectively.
\bibitem {separation} We use $\delta x =\frac{2\hbar k}{M}{(T_{2} - 
T_{1}})-x_{\epsilon}$ where $\frac{2\hbar k}{M}$ = 5.9 cm/s is the two photon
recoil velocity of  sodium and $x_{\epsilon}$ is a small correction
of the order $\delta v \, {T_{1}}$ due to the repulsion 
of the wavepackets.  We include the correction 
in our data analysis in a self-consistent manner.  The correction modifies
$\alpha$ insignificantly, but increases the final values of
$\delta v$ by $\approx$ 0.05 mm/s.
\bibitem{contraction} In fact, after the first pulse the wavepackets 
contract because the reduced mean-field energy can no longer support the wavepacket size.
\bibitem {velcorrection}  We also correct the fringe spacings 
for the contraction of the wavepackets between the final interferometer pulse
and when the image is taken.
\bibitem{approx} We assume $\delta v<<\frac{\hbar k}{M}$, 
$T\omega_{x}<<1$, and harmonic motion in the trap.
\end{references}
\end{document}